\documentclass[12pt]{iopart}
\newcommand{\lone}{${\rm L}_1$}
\newcommand{\ltwo}{${\rm L}_2$}

\usepackage{epsfig}

\begin{document}

\title{Mission design for LISA Pathfinder}
\author{M Landgraf\dag, M Hechler\dag, and S Kemble\ddag}
\address{\dag\ ESA/ESOC, Robert-Bosch-Stra{\ss}e 5, D-64293 Darmstadt, 
Germany}
\ead{Markus.Landgraf@esa.int}

\author{\ddag\ EADS Astrium, Stevenage, United Kingdom}
\submitto{\CQG}

\begin{abstract}
Here we describe the mission design for SMART-2/LISA Pathfinder. The
best trade-off between the requirements of a low-disturbance
environment and communications distance is found to be a
free-insertion Lissajous orbit around the first co-linear Lagrange point of
the Sun-Earth system (\lone), $1.5\times 10^6\:{\rm km}$ from
Earth. In order to transfer SMART-2/LISA Pathfinder from a low Earth
orbit, where it will be placed by a small launcher, the spacecraft
carries out a number of apogee-raise manoeuvres, which ultimatively
place it to a parabolic escape trajectory towards \lone. The challenges
of the design of a small mission are met, fulfilling the very
demanding technology demonstration requirements without creating
excessive requirements on the launch system or the ground segment.
\end{abstract}

\section{Introduction}

ESA has decided to flight-test the technology that its Laser
Interferometer Space Antenna (LISA) will use to detect gravitational
waves (Irion 2002, Danzmann 2003, Hechler and Folkner 2003). The main
goal in this test is to find out whether non-gravitational
disturbances can be removed below the sensitivity limit by careful
design and a drag-free and attitude control system (DFACS). The DFACS
uses an inertial sensor to detect non-gravitational accelerations
acting on the spacecraft hull, and then compensates them using
micro-propulsion. The second mission in the frame of the ESA programme
for Small Missions for Advanced Research in Technology (SMART-2) was
initiated to perform the required space-borne tests. Here we describe
the mission design for SMART-2/LISA Pathfinder. As the interplanetary
target orbit for LISA requires substantial on-board communication
equipment, which is not suitable for a small mission, alternative
options for the operational orbit of SMART-2/LISA Pathfinder are
considered. The driving requirement for the selection of the
operatinal orbit is to find an environment, which does not exceed the
capabilities of the DFACS. This poses mainly requirements on the power
spectrum of spurious accelerations in the frequency band around
$1\:{\rm mHz}$, strength of gravity gradients, thermal variations, and
electro-static charging currents. In combination these requirements
essentially rule out geo-centric orbits.

\section{Selection of Operational Orbit}

The requirements for the operational orbit are:
\begin{enumerate}
\item differential gravity $<2.5\times 10^{-10}\:{\rm m}\:{\rm
s}^{-2}$
\item high thermal stability
\item maximum $\Delta v$ capacity: $3,170\:{\rm m}\:{\rm s}^{-1}$
\item full-year launch window
\item daily visibility from the ESA ground-station in Villafranca
$\geq 8\:{\rm hours}$
\end{enumerate}

\begin{figure}
\begin{center}
\epsfxsize=.5\hsize
\epsfbox[170 440 415 725]{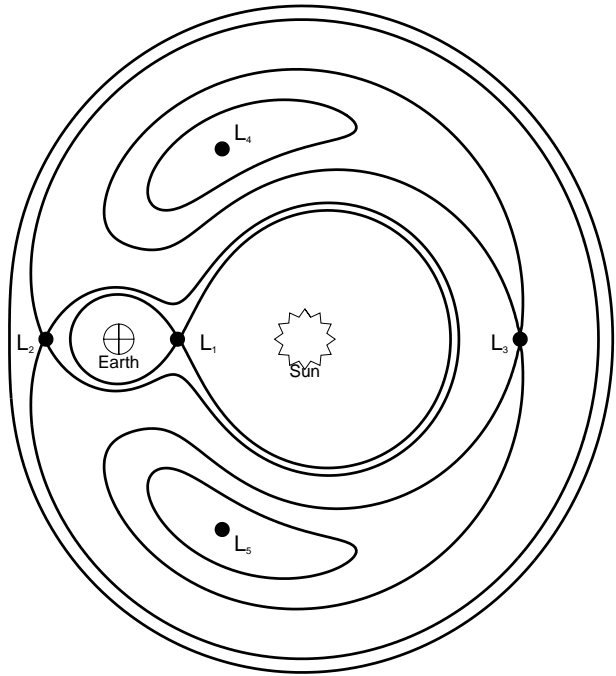}
\end{center}
\caption{\label{fig_zvc} Curves of zero velocity (ZVC) in the RC3P of
Sun, Earth, and spacecraft (not to scale). The near-Earth Lagrange
points \lone and \ltwo are located at the interstection of the ZVCs on
the line connecting Sun and Earth.}
\end{figure}

In order to achieve the required flatness of the gravity field
simultaneously with the thermal stability, an orbit far away from the
Earth must be found without violating the maximum $\Delta v$
capability. Libration orbits around the near-Earth Lagrange points of
the restricted circular three-body problem (RC3P) are located $\approx
1.5\times 10^6\:{\rm km}$ from Earth and can be reached without an
insertion manoeuvre (Farquhar 1973, G{\'o}mez et al 2001).

For LISA Pathfinder the Sun-ward Lagrange point \lone\ is selected in
order to facilitate spacecraft design with the solar panel on one side
and the antenna on the other. The two types of free transfer (no
insertion) libration orbits around \lone\ are distinguished by the
sequence whith which they arrive at the southernmost point, when the
daily coverage from a ground-station at northern geo-centric latitudes
is worst. For type-1 orbits this is the case $102\:{\rm days}$, and
for type-2 orbits $205\:{\rm days}$ after the escape manoeuvre.

\begin{figure}
\begin{center}
\epsfxsize=.45\hsize
\epsfbox{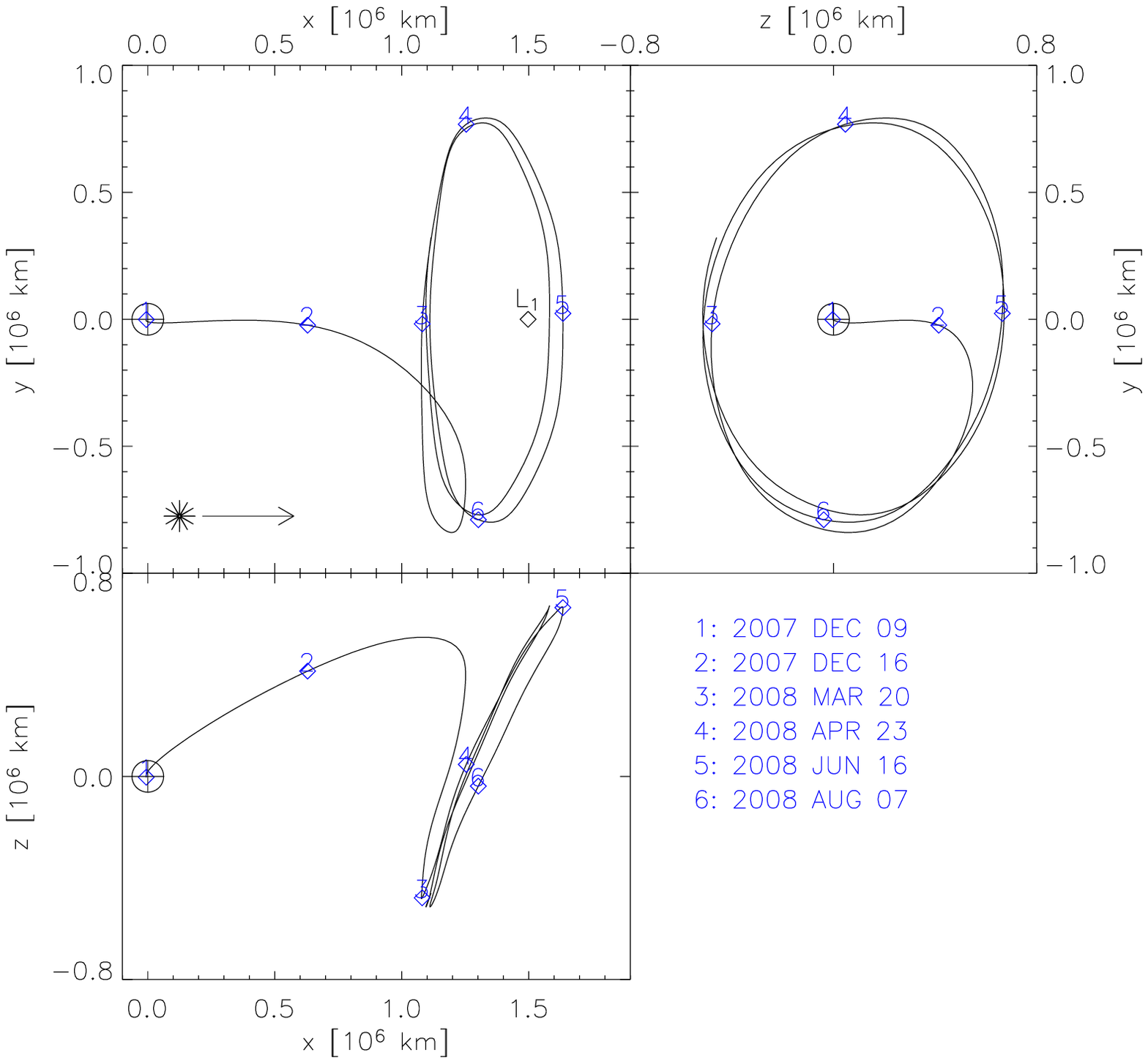}
\epsfxsize=.45\hsize
\epsfbox{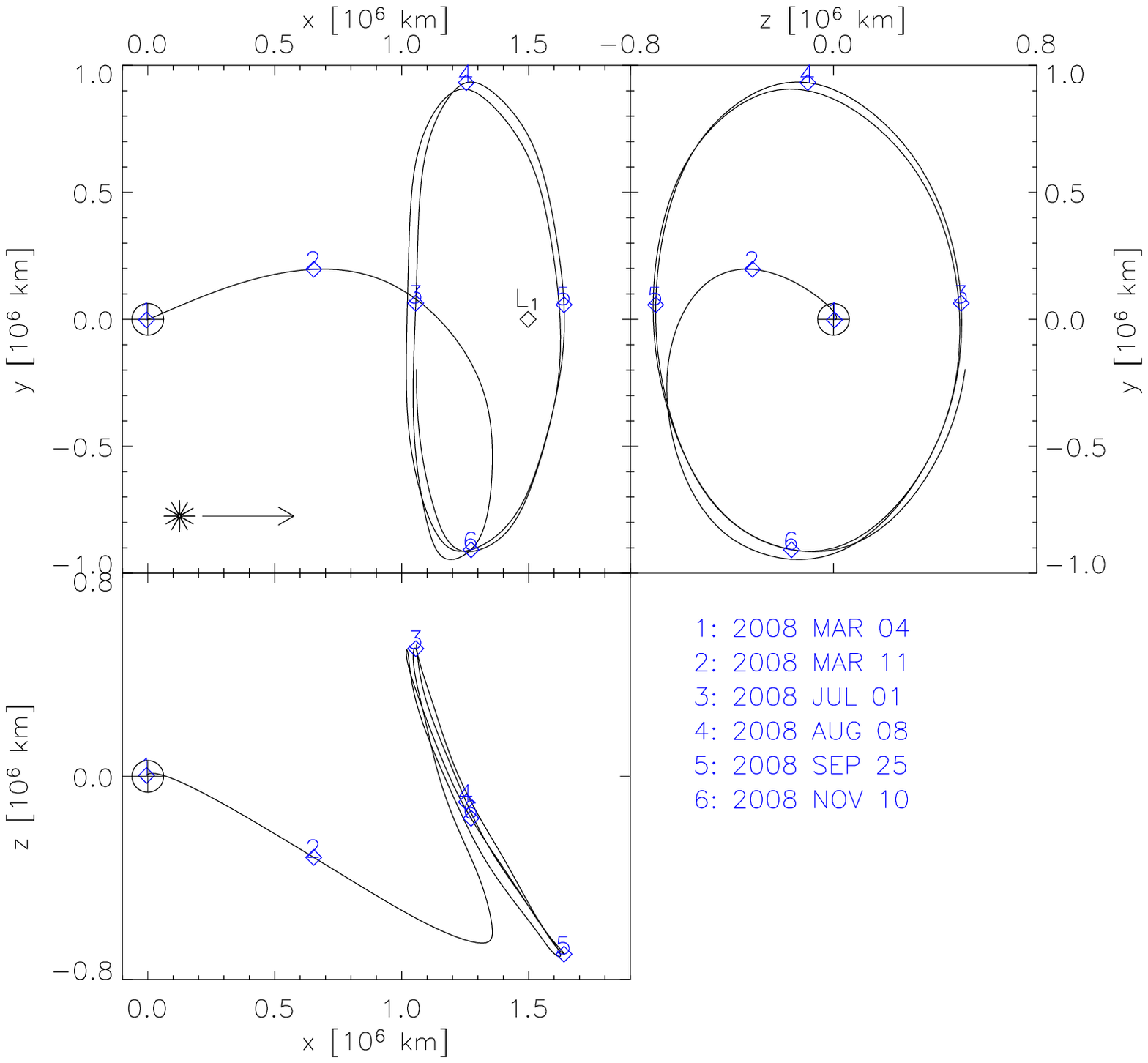}
\end{center}
\caption{\label{fig_reference_ops} Three-dimensional representation of
the type-1 (left) and type-2 (right) trajectories after the escape
manoeuvre. The Earth-centred synodic (rotating) $x$-$y$-$z$ system is
chosen such that the $x$-$y$ plane is the ecliptic plane and the
$x$-axis points towards the Sun.}
\end{figure}

The relatively large libration orbits shown in
figure~\ref{fig_reference_ops} can be reached without insertion
manoeuvre and minimise the variation of the Sun-Earth-spacecraft
angle, which allows to install a fixed horn antenna with a $22^\circ$
beam width ($3\:{\rm dB}$) and thus reasonable gain. The antenna will
be installed such that it points with a cant angle of $20^\circ$ to
$30^\circ$ with respect to the main spacecraft axis, so that the Earth
will always be in the field of view of the antenna if the spacecraft
is rotated around its main axis (which is aligned with the Sun-Earth
line) by $2^\circ$ per day.

\begin{figure}
\begin{center}
\epsfxsize=.9\hsize
\epsfbox{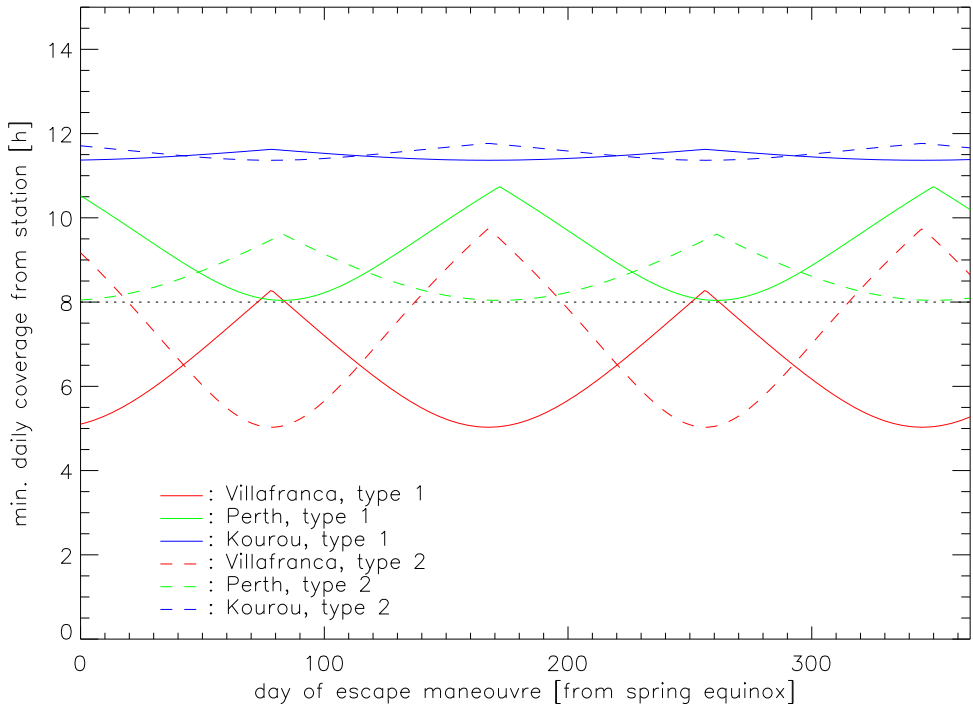}
\caption{\label{fig_vis} Daily visibility of a spacecraft on type-1
and type-2 libration orbits as shown in figure~\ref{fig_reference_ops}
from the ESA ground stations in Villafranca, Perth, and Kourou as a
function of the date of the escape manoeuvre.}
\end{center}
\end{figure}

The downside of these orbits is the large out-of-plane amplitude,
which decreases the daily visibility from a ground-station on the
northern hemisphere of the Earth. As the $15\:{\rm m}$ antenna in
Villafranca ($40^\circ 26'\:{\rm N}$) will be used as the prime
station in the operational phase, type-1 orbits must be selected for
launches around the solstices and type-2 orbits for launches around
the equinoxes. Even when combining type-1 and type-2 orbits, the
launch window does not cover a full year, as illustrated by
figure~\ref{fig_vis}.

\section{Launch and Transfer from Low Earth Orbit}

As LISA Pathfinder will be launched by a small launcher into a
low-Earth orbit (LEO), the total $\Delta v$ to reach the parabolic
transfer is substantial ($>3\:{\rm km}\:{\rm s}^{-1}$). With a small
chemical propulsion module (thrust $400\:{\rm N}$), the escape
manoeuvre must be split in order to avoid an excessive burn duration and 
thus gravity loss. 

\begin{figure}
\begin{center}
\epsfxsize=.5\hsize
\epsfbox{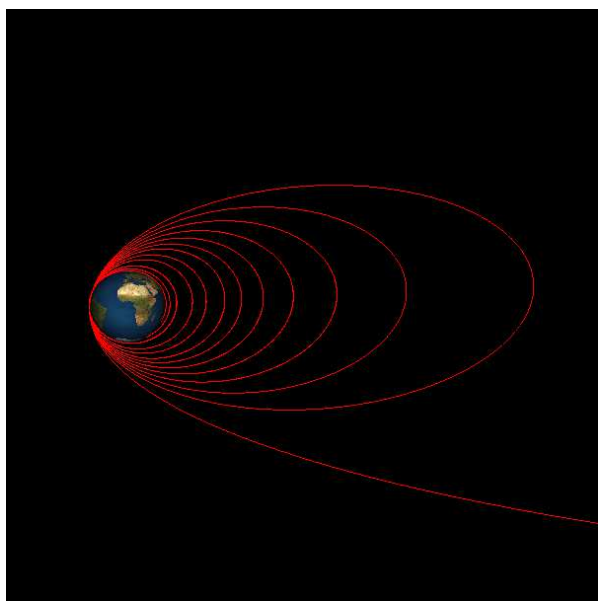}
\end{center}
\caption{\label{fig_sequence_3D} Sequence of orbits after release into
a $900\times 200\:{\rm km}$ LEO by the Launcher. The propulsion module
raises the apogee in $10$ steps to $66,000\:{\rm km}$, before the
eleventh burn puts the spacecraft on an escape trajectory towards
\lone.}
\end{figure}

\begin{figure}
\begin{center}
\epsfxsize=.8\hsize
\epsfbox{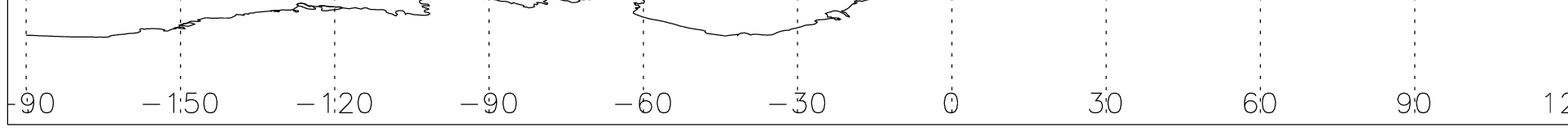}
\end{center}
\caption{\label{fig_groundtrack0} Ground-track of the first 
orbit of the LISA Pathfinder mission. The coverage circles show the
area where the spacecraft is visible to the ground-station in
Kourou (K) and Perth (P). Assuming a launch in
northern Russia, the powered phase of the launch (shown in red) ends
over China. Then, the spacecraft follows a track that leads it to the
injection into the $900\times 200\:{\rm km}$ orbit, which is achieved
by the launcher upper stage (left).}
\end{figure}

\begin{figure}
\begin{center}
\epsfxsize=.8\hsize
\epsfbox{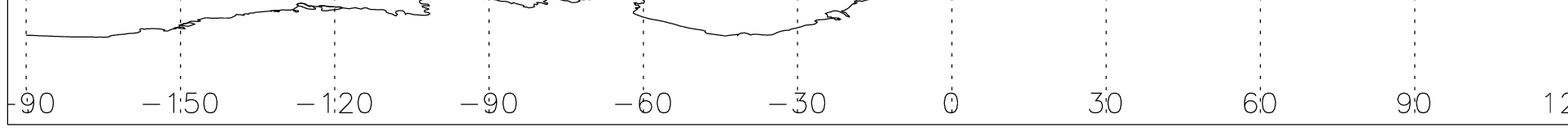}
\end{center}
\caption{\label{fig_groundtrack1} Ground-track of the second and third
orbits of the LISA Pathfinder mission. The coverage circles show the
area where the spacecraft is visible to the ground-station in
Villafranca (V), Kourou (K) and Perth (P). First acquisition of the
spacecraft after injection by a ground-station will be in Perth.}
\end{figure}

For the initial phase of the transfer the visibility by
ground-stations is critical, as shown in
figures~\ref{fig_groundtrack0} and~\ref{fig_groundtrack1}. Only short
passages over the stations occur. The sequence illustrated in figure 4
is optimised so that each burn targets for an apogee pass over a
ground station in Villafranca, Kourou, or Perth. During the pass
ranging and Doppler measurements are taken by analysing the received
signal repeated by the on-board transponder. The accuracy of a single
measurement is $3\:{\rm mm}\:{\rm s}^{-1}$ in Doppler and less than
$1\:{\rm km}$ in ranging. This first pass is to be followed by a
second one, where more orbit determination measurements are
performed. The second pass is needed in order to achieve the required
precision of orbit information before the size of the next manoeuvre
is calculated. The total time allocated for the data processing, orbit
calculation, planning of the next manoeuvre, and preparation of the
commands is 8 hours. After this period, the next pass over one of the
ground stations is used to uplink the commands for manoeuvre execution
during the next perigee. This sequence is repeated for all
apogee-raise manoeuvres during the early transfer phase.

\section{Summary and Discussion}
The challenging mission design for LISA Pathfinder is in an advanced
state, where optimised reference trajectories are established, and
transfer strategies are defined. The environmental requirements set by
the technology demonstration payloads can all be met by placing the
spacecraft in a large amplitude Lissajous orbit around the first
co-linear Lagrange point of the Sun-Earth system. There, the solar
illumination is constant in magnitude as well as direction, minimising
the variability of spurious accelerations and the thermal
equilibrium. 

The transfer sequence puts the spacecraft from an initially low orbit
onto the escape parabola, which leads to a free transfer towards the
large amplitude Lissajous orbit, without violating visibility
constraints from the ESA ground-stations. In theory, the propellant
allocation for the transfer could be reduced by using a lunar swing-by
technique. The saving in $\Delta v$ is however only $50\:{\rm m}\:{\rm
s}^{-1}$, which is negligible compared to the total $\Delta v$ budget
and thus does not justify the increase in operational risk and
complexity.

The current mission design presented above demonstrates how low-cost
missions can be implemented using advanced astrodynamic methods in
ordert to fulfill demanding science requirements under strong
constraints in the total payload mass and launcher performance. With
the LISA Pathfinder mission implemented in this way, the first step
towards detecting gravitational waves in the mHz regime will be taken.

\References
\item[] Danzmann K 2003 {\it Adv. Space Res.} {\bf 32} 1233
\item[] Farquhar R W 1973 {\it Celest. Mech.} {\bf 7} 458
\item[] G{\'o}mez, Jorba {\`A}, Masdemont J, and Sim{\'o} C 2001 {\it
Dynamics and Mission Design Near Libration Points}
\item[] Hechler F and Folkner W M 2003 {\it Adv. Space Res.} {\bf 32} 1277
\item[] Irion R 2002 {\it Science} {\bf 297} 1113
\endrefs

\end{document}